\documentclass{caosp}

\usepackage{graphicx}

\articleNo{146}
\pubyear{2006}
\volume{36}
\volnumber{3}
\firstpage{171}
\lastpage{180}
\received{May 11, 2006}
\accepted{September 6, 2006}

\begin{document}

\hauthor{P.\,Vere\v{s}, L.\,Korno\v{s} and J.\,T\'{o}th}
\title{Search for very close approaching NEAs}
\author{P.\,Vere\v{s}, L.\,Korno\v{s} and J.\,T\'{o}th}
\institute{Department of Astronomy, Physics of the Earth and Meteorology,
Faculty of Mathematics, Physics and Informatics, Comenius University, 842 48
Bratislava, The Slovak Republic}
\date{May 10, 2006}
\maketitle

\begin{abstract}
A simulation of de-biased population of NEAs is presented. The numerical
integration of modeled orbits reveals geometrical conditions of close
approaching NEAs to the Earth. The population with the absolute magnitude up
to $H=28$ is simulated during one year. The probability of possible
discoveries of the objects in the Earth's vicinity is discussed.
\end{abstract}

\section{Introduction}

The population of small NEOs (Near Earth Objects) with the
absolute magnitude up to $H=28$ is still not very well understood.
These objects with the diameter of about 10\,m assuming albedo of
0.14 (Stuart, Binzel 2004) represent transition objects among
asteroids, comets and meteoroids.

As we mentioned in our previous paper (T\'{o}th, Korno\v{s} 2002), discoveries
of new NEOs are influenced by strong observational selection effects. We are
limited by sensitivity of telescopes, a rapid angular velocity at close
encounters in the sky, almost no concentration towards ecliptic - that is why
new NEOs can be found in the entire sky. Current NEO discovery programs (e.g.
Catalina Sky Survey, LINEAR, Spacewatch, NEAT, LONEOS) are primarily focused
on minor planets about 1\,km in diameter and larger. The strategy is to
discover such asteroids in larger geocentric distances, that means to reach
the visual magnitude of about $20-22$, instead of a larger field of view. The
search programs cover the whole sky within one lunation (approx. 20 days). The
magnitude limitation for current NEO discovery strategy exclude smaller
objects in larger distances. These small NEOs could be discovered only during
close encounters with the Earth. But the celestial objects with a rapid
angular velocity are difficult to discover by this type of telescopes. The
current number of known smaller NEOs with $H>19$ decreases rapidly with
increasing absolute magnitude (Fig.\,1).

\begin{figure}
\centerline{\includegraphics[width=5.5cm,angle=-90]{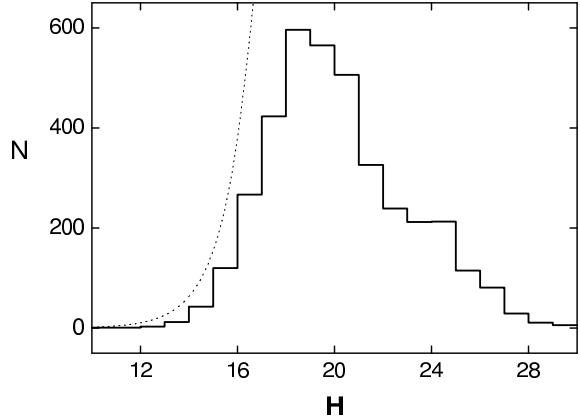}}

\caption{Histogram of the NEO population known by Jan. 5, 2006, versus the
Stuart and Binzel (2004) model (dashed curve).}
\end{figure}

This paper is focused on the population of very small NEOs, up to 10\,m in
diameter and their possible discoveries in the close vicinity of the Earth.
This range size of NEOs is also very dangerous for life on the Earth, as their
atmospheric blasts can produce substantive damage on the surface (e.g.
Tunguska event in 1908). Moreover, the collisions with small objects are much
more frequent than with larger asteroids.

We simulate the orbital distribution of the modeled Near Earth Asteroid
population to estimate the number of close encounters with the Earth within
the mean Earth-Moon (0.0026\,AU) distance. In this close vicinity also a 10\,m
asteroid in suitable geometrical conditions can reach visual magnitude about
$+14^{m}$, which represents the limit for small telescopes with a short focal
length. We also propose a searching program which would be able to discover
such small objects in the close Earth's vicinity. The results from the program
both positive or negative and comparison with our simulated NEAs close
approaches would help to better understand a small asteroid/comets ratio on
the NEO orbits.

\section{The simulation of NEA population}

The current known NEO population is influenced by several observational
effects. Due to this fact, it is necessary to use a debiased model to estimate
the real population and frequency of encounters with the Earth.

We focused our simulation only on NEAs (Near Earth Asteroids) according to
Bottke {\it et al.} (2000), who used three NEA source regions, the $3:1$ mean
motion resonance with Jupiter, the $\nu_{6}$ secular resonance and the
intermediate Mars-crossers source. They produced the NEA debiased orbital and
absolute magnitude size distributions model for $H<18$, calibrated to 910 NEAs
of this size range by fitting it to a biased population of NEO discovered by
Spacewatch.

We extrapolated the de-biased model of the orbital elements distribution (Bottke
{\it et al.}, 2000) of NEA population up to $H=28$, assuming the same delivery
mechanism from the source regions for small NEAs. The orbital elements {\it a, e,
i} follow Bottke {\it et al.} distributions and the angular elements (argument
of perihelion, longitude of ascending node and mean anomaly) were generated
randomly. To calibrate the extrapolated model, we used the NEO cumulative
size distribution model according to Stuart and Binzel (2004)

\begin{equation}
\label{c} N(<H)=10^{-3.88+0.39H}~~,
\end{equation}

\noindent which is a most conservative approximation, compared to
other authors (Bottke {\it et al.}, 2000 or Rabinowitz {\it et
al.}, 1994). A discrete distribution has been used for each
orbital element determination. For our purpose, a small range
(bins) of elements value have been chosen: $\Delta a = 5\times
10^{-6}$ AU, $\Delta e = 10^{-6}$, $\Delta i = 2.5\times 10^{-4}$
deg, $\Delta \omega = \Delta \Omega = \Delta M = 10^{-4}$ deg and
for the absolute magnitude $\Delta H = 10^{-2}$. The NEA $a-e$
phase space condition ($q<1.3$ AU and $Q> 0.983$ AU) was applied
for each generated object. Finally, we got 10\,964\,782 NEA
simulated bodies with the orbital elements and absolute magnitude
distribution (Fig.\,2).

\begin{figure}
\centerline{\includegraphics[width=8.0cm,angle=-90] {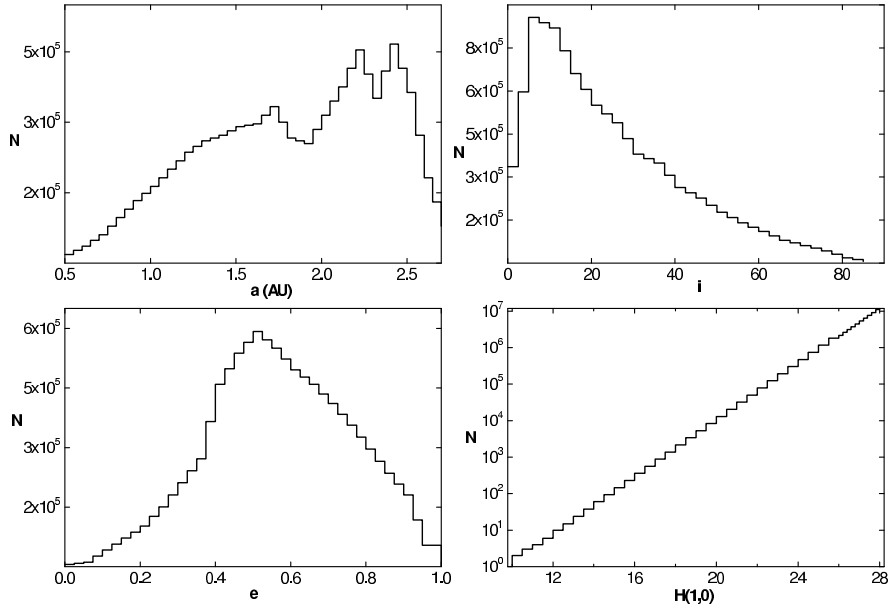}}

\caption{The semimajor axis, inclination, eccentricity and
absolute magnitude (size) distributions of 10\,964\,782 simulated
NEAs.}
\end{figure}

\section{Results}

We numerically integrated 10\,964\,782 modeled NEA orbits during one year with
the aim to study geometrical conditions during their close encounters with the
Earth. We were looking for very close encounters within the Moon distance
(0.0026\,AU) and we were also interested in all asteroids which reached visual
brightness below $+14^{m}$.

In the study for a backward integration of the orbital evolution a DE
multi-step procedure of Adams-Bashforth-Moulton's type up to 12th order, with
a variable step-width, developed by Shampine and Gordon (1975), was used.
Input data of the positions of the Earth, the only considered perturbing body,
were obtained from the Planetary and Lunar Ephemerides DE406 prepared by the
Jet Propulsion Laboratory (Standish, 1998). This simple model is used due to a
short time interval (one year) of integration. Several tests proved that results are
consistent with all major planets incorporated. Each output file contains the
Julian date, the heliocentric and geocentric distances perturbed by the Earth,
the phase angle, the apparent visual magnitude, the right ascension and
declination of the object.

As the main result of the one year simulation, 80 bodies inside
the Moon's orbit were obtained, which implies collisional
frequency with the Earth of $2.10^{-2}$ per year. Only 18 of them
also reached the visual magnitude of $+14^m$. As it could be
expected, the cumulative number of close approaches increases
approximately as a quadratic function of geocentric distance
(Fig.\,3). However, the count of bodies in smaller geocentric
distances is higher than quadratic dependance resulting purely
from Keplerian motion. It is due to the Earth's gravity, although
the effect is minor, still enough to be recognized. Moreover, the
absolute magnitude distribution of the encounters within the
Moon's orbit prefers smaller objects ($<$\,100\,m) with the
maximum at 10\,m, the most numerous objects in the NEA size
distribution (Fig.\,4).

The most important parameter to detect the encountered object by any searching
system is its brightness. The visual magnitude distribution of approaching
objects obtained from our simulation is depicted in Fig.\,5. The majority of
objects within Moon's orbit reached brightnesses between $+16^m$ and $+18^m$.
That is why the limiting magnitude of a small searching system would be the
most restrictive factor of its efficiency.

The next important detection parameter of an object is its angular velocity in
the sky. It represents the effective exposure time during which the object
remains on one pixel of a CCD detector. The angular velocity of the object
depends on its geocentric distance and geocentric velocity vector. The angular
velocities of simulated objects within the Moon's orbit are up to 100
arcmin/min (Fig.\,6), but the most frequent velocities are in the range of
$10-20$ arcmin/min.

\begin{figure}
\centerline{\includegraphics[width=6.5cm,angle=0]{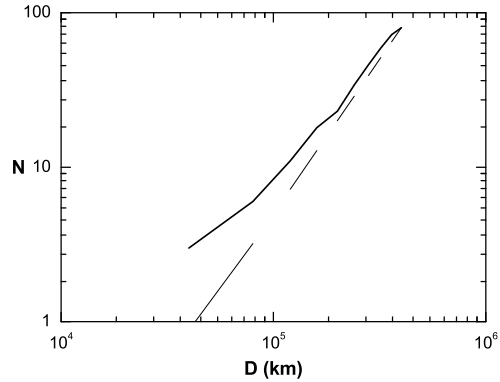}}

\caption{Cumulative annual NEAs flux inside the Moon's orbit (solid),
theoretical quadratic dependance (dashed), D is the minimum geocentric
distance. One simulated object collided with the Earth.}
\end{figure}

\begin{figure}
\centerline{\includegraphics[width=6.5cm,angle=0]{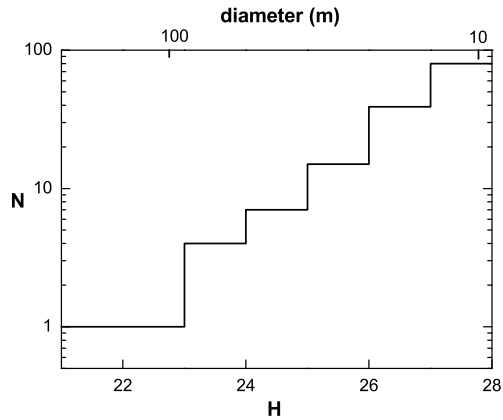}}

\caption{Cumulative annual NEAs flux inside the Moon's orbit. Size
distribution.}
\end{figure}

\begin{figure}
\centerline{\includegraphics[width=5.5cm,angle=-90]{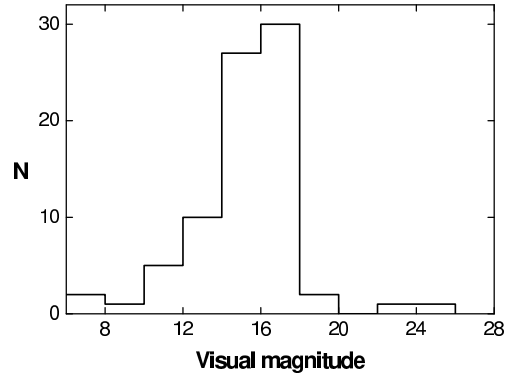}}

\caption{Annual flux of NEAs inside the Moon's orbit and maximum
visual magnitude distribution at the closest encounters.}
\end{figure}

\begin{figure}
\centerline{\includegraphics[width=4.9cm,angle=-90]{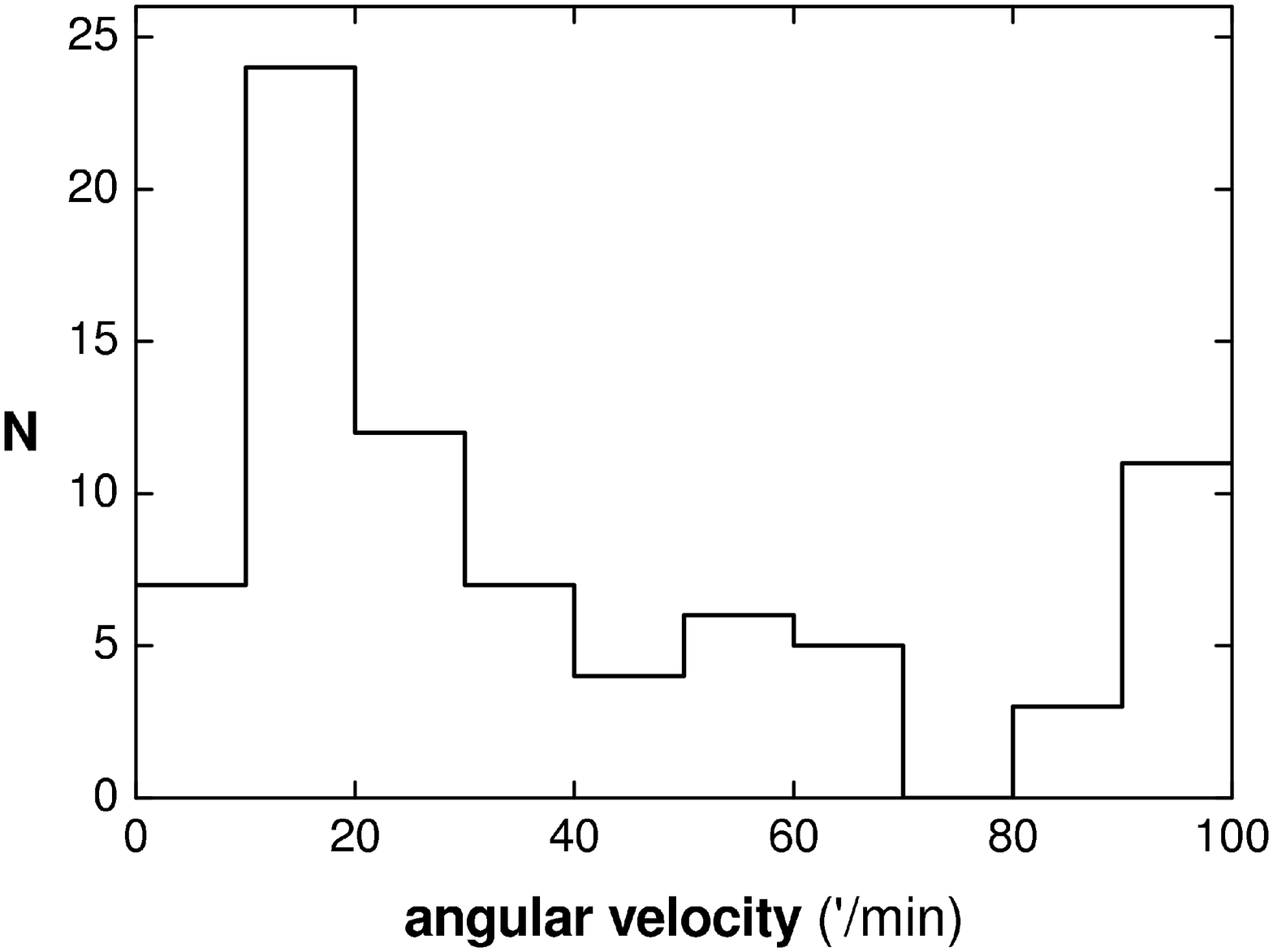}}

\caption{Histogram of angular velocity of NEAs at the closest approaches to
the Earth within the Moon's orbit.}
\end{figure}

The right plot of Fig.\,7 shows all 80 close encounters trajectories within
the Moon's orbit projected on the sky in equatorial coordinates. There is no
noticeable concentration toward the ecliptic. Some objects changed their positions
over 100 degrees in one day. At the same time the phase angles changed very
quickly, which strongly influenced the visual magnitudes of simulated objects.
On the left plot of Fig.\,7 there are depicted only objects inside the
Moon's orbit brighter than $+14^m$.

Also, there was a question how many of simulated objects could be detected by
small telescopes in one year. Totally 119 bodies reached the visual
magnitude of $+14^m$ and brighter. Just 18 of them, as we have already mentioned,
were at the minimum encounter distance within the Moon's orbit. Other bodies,
mostly larger objects, reached $+14^m$ in greater geocentric distances at
various phase angles. There are 52 of them bright enough due to their small
heliocentric distances. There is a possibility to search for such objects at
small elongations from the Sun (mornings and evenings) or by the LASCO
coronograph on SOHO space mission.

\subsection{Possible detections}

We have applied several geometrical conditions and limitations on close encounters
obtained from the simulation in relation to a proposed searching system with the
limiting magnitude of $+14^m$ and weather conditions for Central Europe. In the
estimate we took into account only objects with the absolute magnitude $H>19$,
as we have focused on a search for small, mostly unknown objects, which would
not be discovered by the current searching programs (see Fig.\,1).

\begin{figure}
\centerline{\includegraphics[width=4cm,angle=-90]{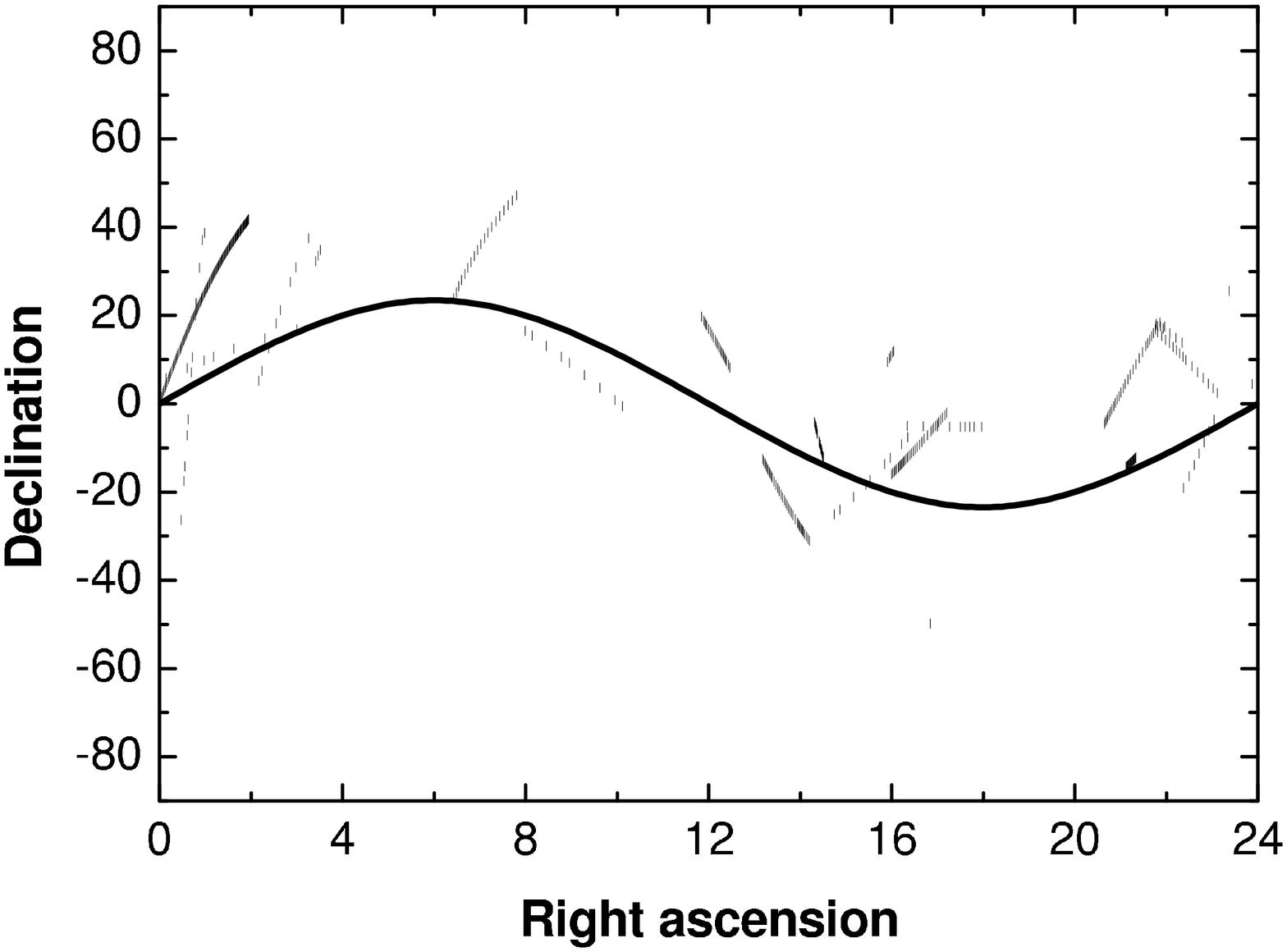}
           \hspace{0.3cm}
            \includegraphics[width=4cm,angle=-90]{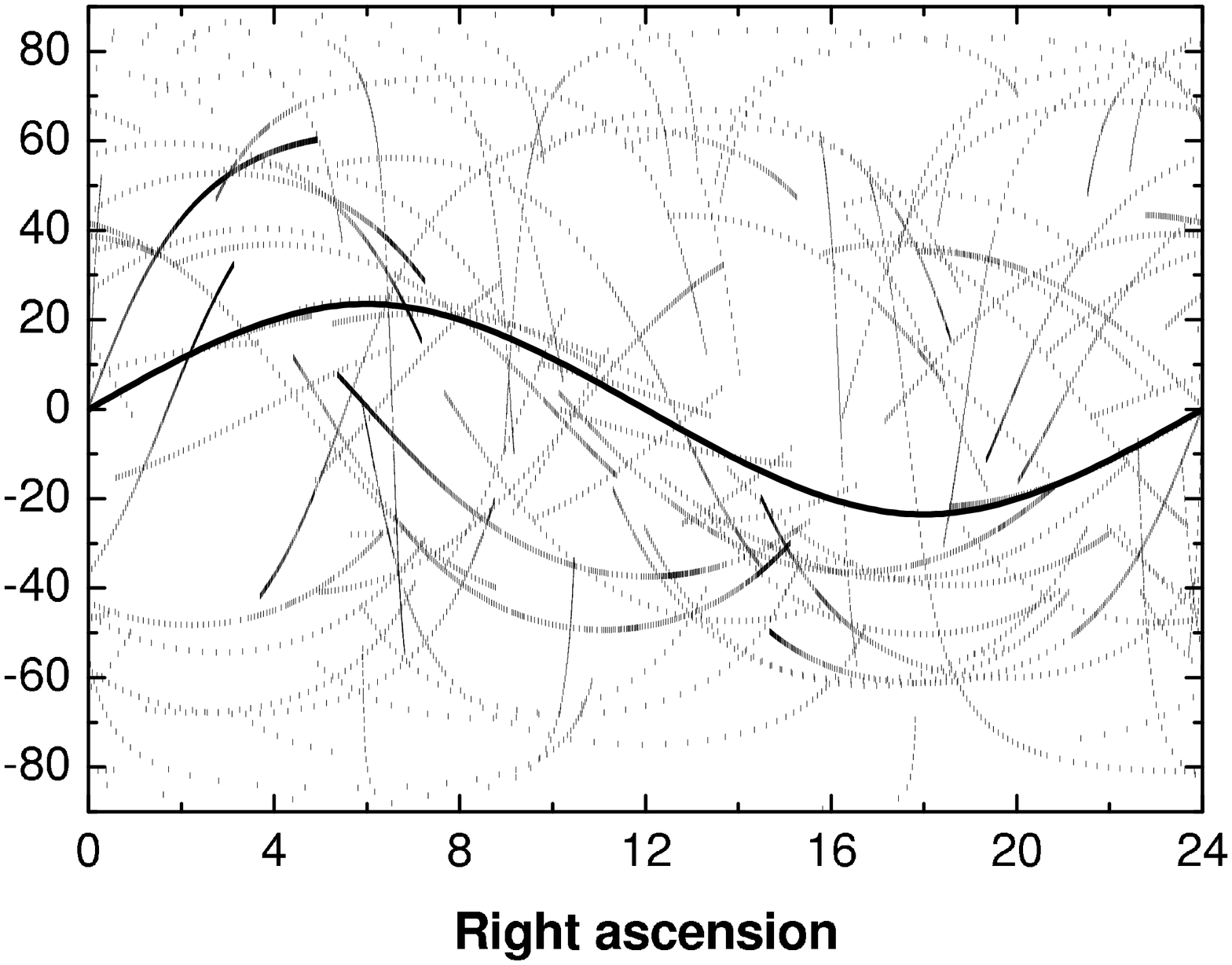}
           }
\caption{Close encounters trajectories within the Moon's orbit projected on
the sky. Simulated objects with the apparent magnitudes $<14^{m}$ (left)
compared to all 80 NEAs inside the Moon's orbit (right). The solid curve is
the ecliptic.}
\end{figure}

The visual magnitude of an object strongly depends on the phase
angle. The brightness at the phase angle $Ph=60\degr$ is about 2
magnitudes lower than that at the opposition point with $Ph=0\degr$
in the same heliocentric and geocentric distances. During a very
close encounter the phase angle of the object is almost equal to
its angle from the opposition point. That is why we suggest an
effective search scan of the sky just within the area of $60\degr$
around the opposition. This part of the sky represents the area of
approximately $10\,000\degr^2$. Due to our upcoming searching
system limitation in the sky coverage area we suppose to cover
only $\sim2\,700\degr^2$, which corresponds to the phase angle
$Ph=30\degr$.

The effective time which the object spends in the searching area ($30\degr$
from the opposition) reduces the probability of the detection. Another
decrease occurs after reducing the magnitude of each object according to its
angular velocity, which corresponds to the effective exposure time per pixel
for a particular optical CCD system. There is also taken into account the time
of the encounter that have to be in the nighttime for Central Europe.

\begin{table}[h]
\small
\begin{center}
\caption{The simulation results based on the NEAs ($H<28$) population estimate
by Stuart and Binzel (2004). The number of NEAs per year: (a) brighter than
$+14^{m}$ with no distance limitation; (b) inside the Moon's orbit; (c) inside
the Moon's orbit and brighter than $+14^{m}$; (d) brighter than $+14^{m}$,
observable in Central Europe and $H>19$; (e) possible discoveries by the
proposed searching system. Presumptive results are based on the population
estimates by Bottke {\it et al.} (2000) and Rabinowitz {\it et al.} (1994).}

\begin{tabular}{lccc}
\hline\hline N annually & Simulation  & Bottke {\it et al.}, 2000 & Rabinowitz
{\it et al.}, 1994 \\ \hline a) apparent mag.$<14$ & 119 & 725 &
2636\\ b) inside Moon orbit & 80 &528 &1920\\ c) inside Moon orbit and \\
~~~~vis. mag.$<14$ & 18 & 118 & 432\\ d) vis. mag.$<14$ and & & &
\\ ~~~~observable in CE, $H>19$ & 11& 66& 264\\ e) poss. discoveries
by s.s. & 3 & 18 & 72\\ \hline \hline
\end{tabular}
\end{center}
\end{table}

According to these conditions we obtained 18 possible detections during one
year simulation for the Modra observatory (Comenius University, Bratislava,
Slovakia). Only 6 of them occurred inside the Moon's orbit. One simulated body
hit the Earth. The number of bodies in suitable geometrical conditions
slightly varies with a location on the Earth. The last reduction is needed due
the fact that only about $20-30$ percent of the nights in Central Europe are
in good weather conditions.

Finally, we obtained only $3-5$ objects, which would be detected per year with
the proposed searching system for the particular weather conditions in Central
Europe. This result is not very optimistic, but we have chosen for our
simulation the conservative estimate of the NEA population ($H<28$) by Stuart
and Binzel (2004). The total NEO population estimates up to $H<28$ by other
authors like Bottke {\it et al.} (2000) or Rabinowitz {\it et al.} (1994) are
several times ($6-24$ times) larger (Tab.\,1).

Especially, analysis of objects inside the Moon orbit showed that a higher
limiting magnitude of a wide field survey system leads to more discoveries. As it
is clear from the visual magnitude distribution in Fig.\,5, the $+14^{m}$
limiting magnitude is not sufficient for the effectivity of the searching
system. To detect a majority of close encounters the $+18^m$ magnitude would
be needed.

\subsection{Searching system}

The simulation of geometrical conditions during close approaches to the Earth
were done with the aim of possible discoveries of small asteroids in the
Earth's vicinity.

An example of such a searching system is described below
(Fig.\,8). The parameters of the system with the CCD ST8 camera
are as follows: the focal length $f=150$\,mm, the aperture
$D=180$\,mm, the resolution of 37.15 arcsec per pixel, the field
of view $5.27\degr\times 3.51\degr$, the exposure time of 30\,s,
which corresponds to the limiting magnitude of $+13.2^{m}$.

\begin{figure}
\centerline{\includegraphics[width=8cm,angle=0]{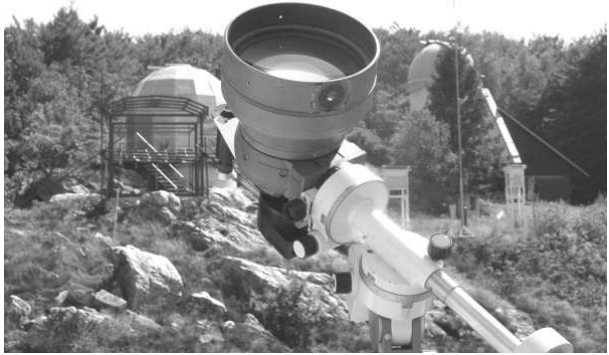}}

\caption{An example of a possible searching system for the close approaching
NEOs.}
\end{figure}

The pro/-posed wide-/-field short focal length search telescope is not suitable
for precise astrometric observations, but is still useful to follow up detection
of close approaches. Such observation can provide a long arc with the sufficient
curvature for preliminary orbit determination (Milani, 2006), afterwards
confirmed by our f/5.5 60\,cm reflector. This kind of detection could be
classified as the original discovery.

The searching system is not finished yet and is in an early testing phase. But
the most restrictive factor of such a short focal length system is a low
limiting magnitude. The results of our simulation showed that the effective
searching system has to have the limiting magnitude up to $+18^{m}$ and would
cover not less than 10\,000$\degr^2$ per night scan rate (Fig.\,5).

\section{Conclusions}

We have extrapolated the de-biased model of NEA orbital distribution (Bottke {\it
et al.}, 2000) up to $H=28$ and calibrated it using the NEO cumulative size
distribution model of Stuart and Binzel (2004). We have numerically studied of about 11
millions generated NEAs during one year. As the main result from the
simulation we obtained 80 close encounters with the Earth within the Moon's orbit.
Only 18 of these objects reached the visual magnitude of $+14^{m}$.

The pro\-posed wide-field short focal length survey system, assuming a
ground-\-based restrictions, observational se\-lection effects and atmo\-spherical
conditions could find several ($\sim 3$) NEAs per year. If the real popu\-lation
is larger (Rabinowitz {\it et al.}, 1994; Bottke {\it et al.}, 2000), there is
a possibi\-lity to find $18-72$ objects per year. An in\-crease of the limiting
magnitude of the survey system to $+18^{m}$ with pre\-serving a wide field of
view could lead to more disco\-veries.

An alternative way to compare our simulation results with
observations is to calculate an Earth's annual collisional
frequency. The collisional frequency of 10\,m objects calculated
from DoD satellite (U.S. Department of Defence) observations of
large bolides is $10^{-1}$ (Brown {\it et al.}, 2002). Our
simulation resulted in the collisional frequency of $2.10^{-2}$ per year,
which implies that the real population should be at least 5 times
more numerous than that used in our expanded model up to $H=28$
based on Stuart and Binzel (2004).

\acknowledgements The authors thank to P. Brown and A. Gal\'{a}d for valuable
comments. This work was supported by the Slovak Grant Agency, Grant No.
1/3067/06.

\end{document}